\documentclass[aps,twocolumn,prl,showpacs]{revtex4}
\usepackage{amsmath,amssymb,latexsym,graphicx}

\begin{document}
\title{Ramsauer-Townsend Diffraction Oscillations in the Two-Dimensional Momentum Distribution of Laser-Ionized Electrons}
\author{D. G. Arb\'{o}} 
\affiliation{Institute for Theoretical Physics, 
Vienna University of Technology, Wiedner Hauptstra\ss e 8-10/136, A-1040 Vienna, Austria}
\author{S. Yoshida}
\affiliation{Institute for Theoretical Physics, 
Vienna University of Technology, Wiedner Hauptstra\ss e 8-10/136, A-1040 Vienna, Austria}
\author{E. Persson}
\affiliation{Institute for Theoretical Physics, 
Vienna University of Technology, Wiedner Hauptstra\ss e 8-10/136, A-1040 Vienna, Austria}
\author{K. I. Dimitriou}
\affiliation{Institute for Theoretical Physics, 
Vienna University of Technology, Wiedner Hauptstra\ss e 8-10/136, A-1040 Vienna, Austria}
\author{J. Burgd\"{o}rfer}
\affiliation{Institute for Theoretical Physics, 
Vienna University of Technology, Wiedner Hauptstra\ss e 8-10/136, A-1040 Vienna, Austria}

\date{\today}
\begin{abstract}
We analyze the two-dimensional momentum distribution of electrons ionized by few-cycle laser pulses in the transition regime from multiphoton absorption to tunneling by solving the time-dependent Schr\"odinger equation and by a classical-trajectory Monte Carlo simulation with tunneling (CTMC-T). We find a complex two-dimensional interference pattern that resembles ATI rings at higher energies and displays Ramsauer-Townsend diffraction oscillations in the angular distribution near threshold. CTMC-T calculations provide a semiclassical explanation for the dominance of selected partial waves. While the present calculation pertains to hydrogen, we find surprising qualitative agreement with recent experimental data for rare gases \cite{Rudenko04}.
\end{abstract}
\maketitle
 The interaction of few-cycle laser pulses with matter has recently attracted considerable interest \cite{Paulus01} as increasingly shorter pulses with duration of the order of $10$ fs and below became available. Novel aspects of laser-matter interactions such as the dependence of high-harmonic radiation or electron emission  on the carrier envelope phase \cite{HHG,electron} and the interference of electronic wave packets emitted at different points in time during the ultrashort pulse \cite{lind} became apparent. Another recent advance is the imaging of the momentum distribution of the ionized electron providing insight into the ejection of both one-electron \cite{Rudenko04} and non-sequential multiple electron emission \cite{doerner}. For single-electron emission, the longitudinal  momentum distribution ($k_z$ along the direction of the laser polarization) of photoelectrons from rare gases features a broad ``double-hump'' structure near threshold which surprisingly resembles the $k_z$ distribution for non-sequential double ionization \cite{Moshammer03}. While for the latter case this structure results from electron-electron collision during rescattering of the laser driven electron at the ionic core, in the former case it is due to the interplay of the Coulomb interaction and laser field on the receding trajectory \cite{Moshammer03,chen,Dimitriou04,Faisal05}. By contrast, the transverse momentum distribution features a narrow Coulomb-like cusp well-known from ion-atom collisions \cite{Dimitriou04,Rudenko05,Comtois05}. Very recently, Rudenko et al. \cite{Rudenko04} presented first fully two-dimensional $(k_z, k_\rho)$ momentum maps for laser-ionized electrons from different rare gases, displaying a complex pattern whose origin is, so far, unexplained. In this letter we investigate the 2D momentum map for laser ionization of hydrogen and find an equally complex yet surprisingly similar pattern suggesting a simple explanation independent from specific properties of the target atom. 

The interaction between the laser field and the atom can be characterized by two different mechanisms controlled by 
the value of the Keldysh parameter $\gamma =\sqrt{I_{p}/2U_{p}},$ where $%
I_{p}$ is the ionization potential of the atom, $U_{p}=$ $F_{0}^{2}/4\omega
^{2}$ the ponderomotive energy, $\omega $ the laser angular frequency, and $F_{0}$
the peak amplitude of the laser field. In the multi-photon regime $(\gamma \gg 1)$ the atomic interaction is governed by the quantum nature of the radiation field resulting in the absorption of $n$ photons $(n \geq 1)$ from the field. By contrast, in the tunneling regime $(\gamma \ll 1)$ the atom responds to the strong perturbation by a  ``classical'' electric field where ionization proceeds via tunneling. The present calculation as well as recent experiments \cite{Rudenko04} explores the transition regime around $\gamma \approx 1$ where a more complex response is to be expected. We focus on 
hydrogen in order to avoid any ambiguity resulting from additional
approximations required for many-electron targets. 

The Hamiltonian of a
hydrogen atom driven by a linearly polarized laser field is 
\begin{equation}
H=\frac{{\vec{p}\,}^{2}}{2}+V(r)+z\,F\,(t),  \label{hami}
\end{equation}
where $V(r)=-1/r$ is the Coulomb potential energy, $\vec{p}$ and $\vec{r}$
are the momentum and position of the electron, respectively, and $F(t)$ is
the time dependent external field linearly polarized along the $\hat{z}$
direction. The laser pulse is chosen to be of the form 
\begin{equation}
F(t)=F_{0}\sin ^{2}\left( \frac{\pi t}{\tau }\right) \cos (\omega t+\varphi
)\quad (0\leq t\leq \tau )\;,  \label{field}
\end{equation}
where $\omega $ is the laser frequency, $\varphi = 0$ the
carrier-envelope phase, $\tau $ the total pulse duration, and $F_{0}$ is the
peak field. 
Atomic units are used throughout.

The time-dependent Schr\"{o}dinger equation (TDSE) can be solved by different techniques \cite{Dionissopoulou97,Wassaf03}. Approximation methods include semiclassical approximation
methods \cite{Sand00,Milosevic03}, the (Coulomb-)Volkov approximation 
\cite{Macri03,Rodriguez04} and CTMC-T method \cite{Dimitriou04,cohen}. We employ the
generalized pseudo-spectral method for solving the TDSE~\cite{tong97}. The method combines a
discretization of the radial coordinate optimized for the Coulomb
singularity with quadrature methods to allow stable long-time evolution
using a split-operator method. Both the unbound as well as the bound parts of the wave function $\left| \psi (t)\right\rangle $ can be accurately represented. The calculation of the 2D momentum distribution requires projection of the partial waves $\left| k,l \right\rangle$ onto outgoing Coulomb waves 
\begin{equation}
\frac{dP}{d\vec{k}}=\frac{1}{4\pi k}\left| \sum_{l}e^{i\delta _{l}(k)}\ 
\sqrt{2l+1}P_{l}(\cos \theta _{k})\ \left\langle k,l\right. \left| \psi
(\tau )\right\rangle \right| ^{2},  \label{coulomb}
\end{equation}
after the conclusion of the pulse. In Eq. (\ref{coulomb}) $\delta _{l}(k)$ is the momentum-dependent Coulomb phase shift, $%
\theta _{k}$ is the angle between $\vec{k}$ and the polarization direction
of the laser field, $\widehat{z}$, $P_{l}$ is the Legendre
polynomial of degree $l$, and $\left| k,l\right\rangle $ is the eigenstate of the atomic
Hamiltonian with positive eigenenergy $E=k^{2}/2$ and orbital quantum number 
$l$. The atom is initially in its ground state. Due to the cylindrical symmetry of our system for a linearly polarized laser field, $m = 0$ is a constant of motion and omitted. The distortion of the momentum distribution due to long-range final-state Coulomb interactions is fully accounted for Eq.~(\ref{coulomb}). We also have performed classical-trajectory Monte-Carlo calculations \cite{Dimitriou04} incorporating tunneling (CTMC-T) which include both Coulomb and laser field interaction non-perturbatively.

Examples of the two-dimensional momentum distribution $(k_\rho, k_z)$, $\frac{d^2 P}{dk_\rho dk_z} = 2 \pi k_\rho \left( \frac{dP}{d \vec{k}} \right), $ for an 8-cycle pulse $(\tau = 1005)$, frequency $\omega = 0.05$ and different field amplitudes  $F_0 = 0.0377 \,( \gamma = 1.34), 0.0533 \, (\gamma = 0.95), $ and $0.075 \, (\gamma = 0.67) $ are shown in Fig. 1, illustrating the transition from the multi-photon to the tunneling regime. Each frame displays a complex interference pattern which is characterized by a transition from a ring-shaped pattern at larger $k = \sqrt{k_\rho^2 + k_z^2} \gtrsim 0.4$ with circular nodal lines to a very different pattern of pronounced radial nodal lines for small $k$ near threshold. The first point to be noted is that the overall pattern displays a surprising and striking similarity to the experimental pattern observed recently for rare gases, such as helium, neon, and argon \cite{Rudenko04}. The ring pattern is reminiscent of ATI peaks of the multi-photon regime. The point to be noticed is that ATI rings are present well into the tunneling regime for larger energies $E = \frac{1}{2} k^2 \gtrsim 1 $. Concurrent CTMC-T calculations for the present laser parameters show that classical rescattering provides only a minor contribution to the electron spectra above $E = 2 U_p$. Energies $E\gtrsim 2 U_p$ are effectively inaccessible by the classical quiver motion thus rendering the multiphoton quantum process as dominant pathway.  The ATI-like component can be quantified by the partial ionization probability $p_l^i$ with orbital quantum number $l$ residing within a given $i$-th ring between adjacent minima $(k_i - \Delta_i, k_i + \Delta_i)$
\begin{equation}
\label{eq:4}
p^i_l = \int \limits^{k_i + \Delta_i}_{k_i - \Delta_i} kdk | \langle k,l|\psi (\tau) \rangle|^2 \, .
\end{equation}
The $i$-th circle has a mean radius $k_i = \sqrt{2 E_i}$, where the energy $E_i$ corresponds to the $i$-th ATI peak of the photoelectron spectrum. For $i \gtrsim 2$ $(k > 0.4)$ rings become recognizable.

\includegraphics[width=6cm,angle=270,bb=45 50 570 775]{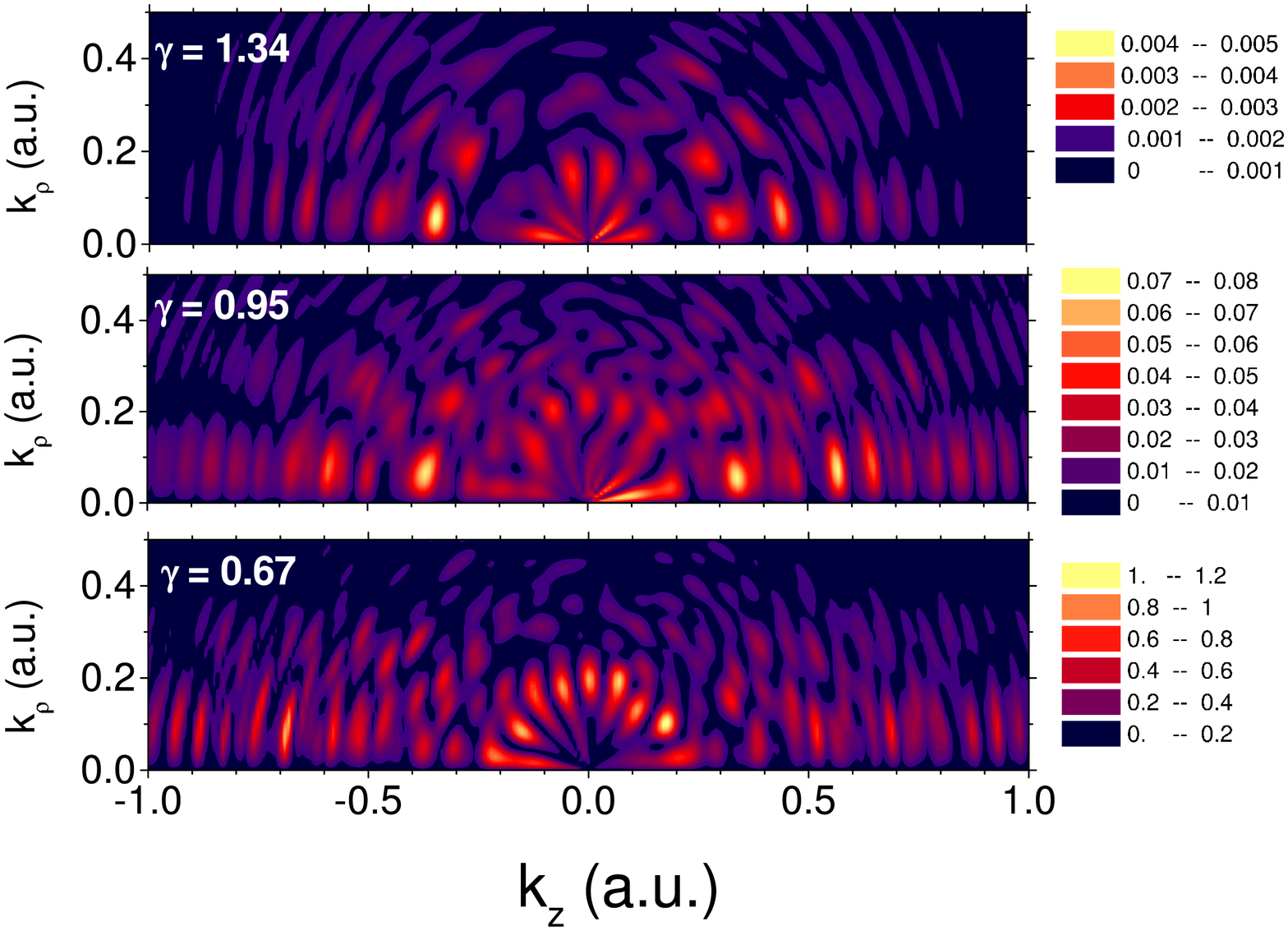}\\
{\bf FIG. 1}: Doubly-differential electron momentum distributions in
cylindrical coordinates $(k_{z}, k_\rho)$. The parameters of the field
are $\omega =0.05$, $\tau =1005$. In (a) $\gamma =1.34$ ($F_{0}=0.0377$),
(b) $\gamma =0.95$ ($F_{0}=0.053$), and (c) $\gamma =0.67$ ($F_{0}=0.075$).\\

The transition to an entirely different and unexpected radial pattern occurs for energies within $ \approx 2 \omega$ from the threshold. The radial nodal pattern at low energies can be made more explicit by analyzing the angular differential probability, $d^2 P / d k d (\cos \theta_k$), at fixed $k$ (Fig. 2). The probability displays pronounced oscillations that remarkably resemble those of a single Legendre polynomial,
\begin{equation}
\label{eq:5}
\frac{d^2 P}{d k d (\cos \theta_k)} \approx \left[P_{l_0} (\cos \theta_k) \right]^2 \, .
\end{equation}  
At lower field and higher $ k = 0.34$, the Legendre polynomial with $l_0 = 6$ dominates while at higher field and low $k = 0.19$ the polynomial $l_0 = 8$ has the largest weight. The dominance of a single Legendre polynomial $P_l$ implies the dominance of a single or a few partial waves in the momentum-differential ionization cross section. The resulting radial nodal pattern with pronounced minima at certain angles is a well-known feature in low-to-intermediate energy electron-atom scattering referred to as generalized Ramsauer-Townsend (GRT) diffraction oscillations \cite{barton,egel,burg}. The present result suggests that laser-driven scattering of electrons in the Coulomb field of the nucleus leads to similar Ramsauer-Townsend-like interference fringes in the angular distribution. The distribution of contributing partial waves, $p_l$, is presented in Fig.~3. Marginally into the multi-photon regime $(\gamma = 1.34, $ Fig.~3a), the $l$ distribution of the first ring displays a broad distribution peaking at odd angular momenta $l = 3, 5, $ and $7$ while in the second ring the distribution features a single peak at $l_0 = 6$ in agreement with $d^2 P/dk d(\cos \theta_k)$ (Fig.~2a). In the tunneling regime ($\gamma = 0.67$, Fig.~3b), the first ring near threshold peaks at $l_0 = 8$. The dominance of a single $l$ is further enhanced by the relative suppression of adjacent angular momentum $l_0 \pm 1$ of opposite parity, which is a remnant of the multi-photon parity selection rule. The second ring shows a peak at $l_0 = 9$, even though no dominance of a single partial wave is evident when looking at the angular distribution (not shown).

\includegraphics[width=7cm,angle=0,bb=20 60 560 770]{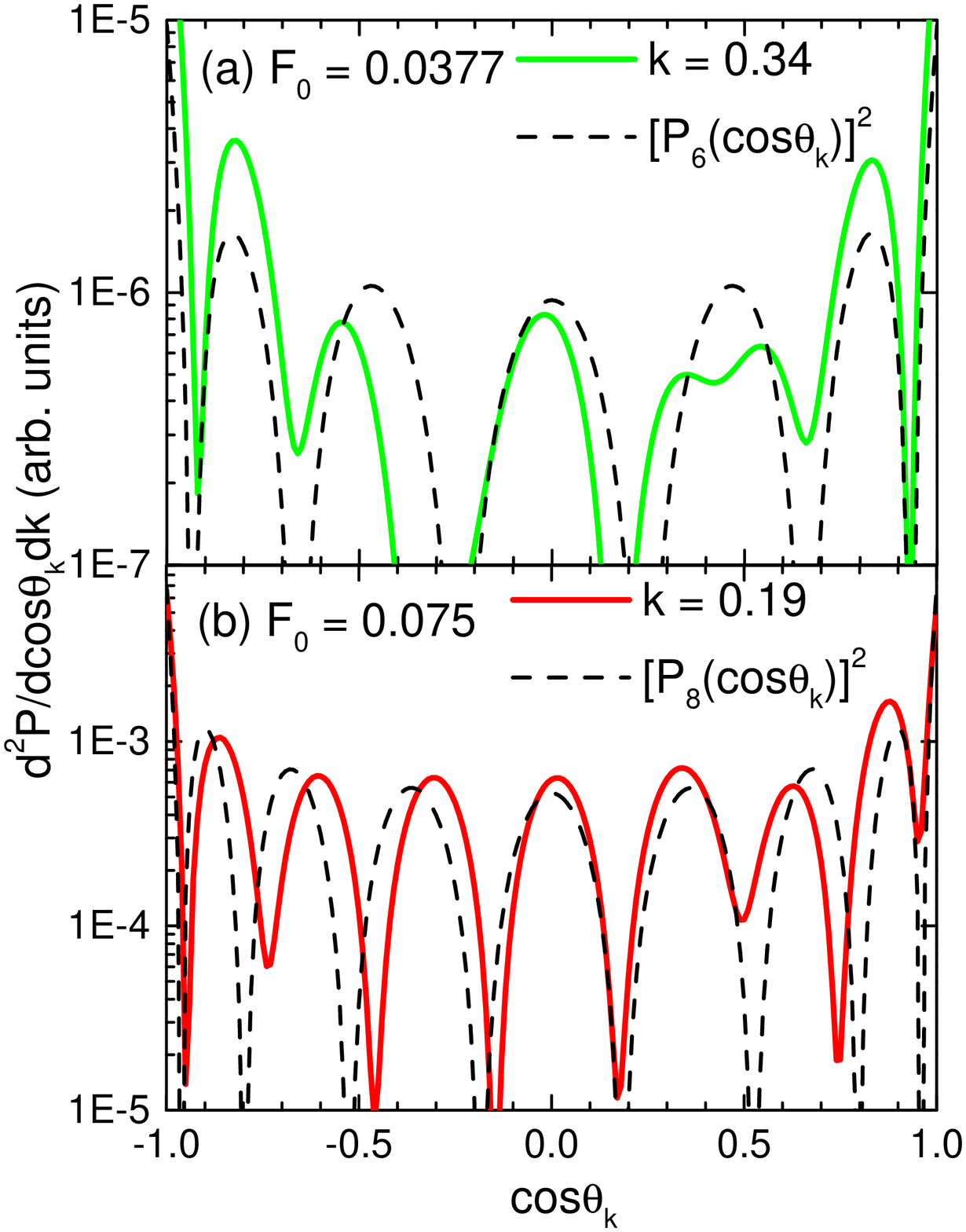}\\
{\bf FIG. 2}: Angular distribution calculated by solving the TDSE (thick solid lines) for $\omega =0.05$, $\tau =1005$. In (a) $\gamma =1.34$ ($F_{0}=0.037$) and $k=0.34$, and in (b) $\gamma =0.67$ ($F_{0}=0.075$) and $k=0.19$. Dashed lines: square of the Legendre 
polynomial $P_{l_0} (\cos \theta_k), l_0 = 6$ in (a), and $ l_0 = 8$ in (b).\\

GRT interference fringes in electron-atom scattering can be semiclassically described in terms of interferences of paths  with (in general) different angular momenta scattered into the same angle $\theta_k$ \cite{burg}. In the special case that a single partial wave and thus a single Legendre polynomial $P_{l_0} (\cos \theta_k)$ dominates, the angular distribution of the interfering paths must have very similar classical ``impact parameters'' such that they belong to the same angular momentum quantum bin $[l_0, l_0 + 1]$ \cite{burg}. An analogous path interference occurs in laser-atom ionization in the tunneling regime near threshold. To uncover the relevant classical paths we employ a CTMC-T simulation \cite{Dimitriou04} for the same parameters as in Fig.~3b. The ensemble of ionized electrons near threshold features, indeed, an $l$ distribution (Fig.~4a) that resembles the quantum distribution (Fig.~3b) with a peak near $l_0 = 8$; clearly emphasizing the underlying classical character of this process. A typical electron trajectory after tunneling shows a quiver motion along the polarization of the laser field. An interesting observation is that, even though the motion is strongly driven by the laser field, the motion follows the Kepler hyperbola (Fig.~4b). The point to be emphasized is that the dashed line in the figure does not represent the laser-driven trajectory averaged over a quiver oscillation period but the unperturbed Kepler hyperbola with the identical asymptotic momentum as the laser driven trajectory (solid line). Thus, the angular momentum of the Kepler hyperbola is identical to that of the asymptotic $l$ of the laser-driven electron. Note that the number of quiver oscillations along the Kepler orbit is not unique thus allowing for path interferences. The distance between the hyperbola and the nucleus at the pericenter is given by \cite{landau} $r_{\min} = \left( \sqrt{1+(k l)^2} - 1 \right) / k^2$.	This distance should be equal to the quiver amplitude $\alpha = F_0 / \omega^2, $ i.e. $ r_{\min} \cong \alpha$. The initial conditions for the laser driven trajectory are provided by tunneling ionization with the release of the electron with zero longitudinal velocity at times $t_i$ near the maxima of the field amplitude $F(t_i) \simeq F_0$. Trajectories released at different times $t_i$  or different maxima of the field reaching the same asymptotic branch of the Kepler hyperbola will interfere and generate GRT fringes. In order to reach the limiting case of the dominance in the semiclassical domain of a single $P_{l_0}$ it is necessary that interference trajectories at fixed energy exist that approximately cover the entire range of scattering angles $(0 \lesssim \theta \lesssim \pi)$ all of which with angular momenta close to $l_0$. Our CTMC-T calculations show that close to threshold $(k \lesssim 0.2)$ such families of trajectories indeed exist. They correspond to Kepler hyperbolas with very similar opening angles of their asymptotes but with the angle of major axis  relative to the laser polarization, $\theta$, distributed between $0 \lesssim \theta \lesssim \pi$. 

\includegraphics[width=7cm,angle=0,bb=20 250 550 760]{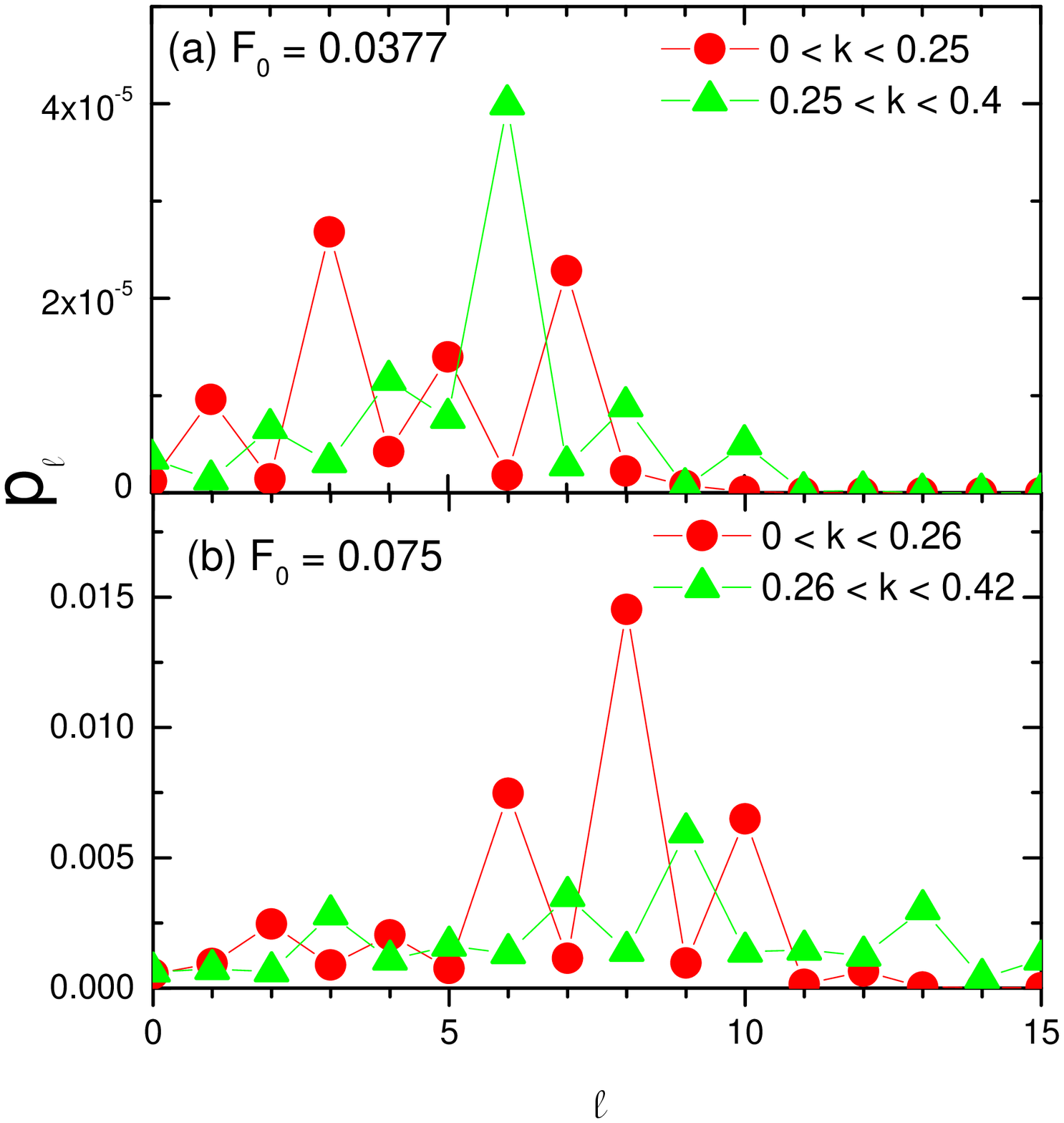}\\

{\bf FIG. 3}: Partial ionization probability, $p_{l}$, as a function of the
angular momentum $l$ for different spectral regions indicated in the figure. The
parameters of the field are $\omega =0.05$, $\tau =1005$, (a) $\gamma =1.34$
($F_{0}=0.0377$) and (b) $\gamma =0.67$ ($F_{0}=0.075$).

\includegraphics[width=8cm,angle=0,bb=50 190 500 680]{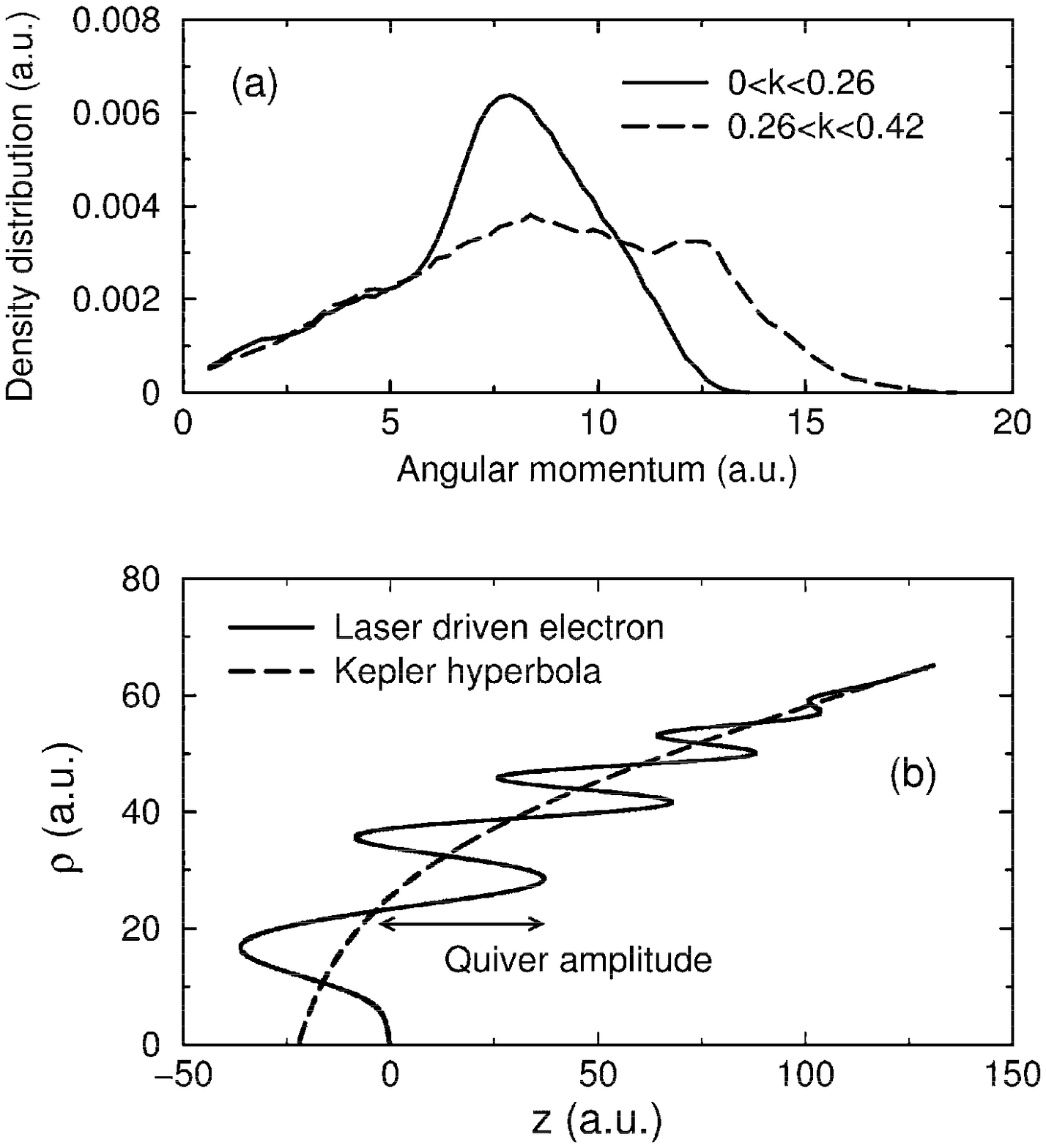}\\
{\bf FIG. 4}: (a) $l$ distribution of classical trajectories for the same parameters as in Fig. 3 (b). (b) Classical trajectory of laser-driven electron (solid line) and unperturbed Kepler hyperbola of same asymptotic $E$ and $l$ (dashed lines). Interferences occur for emission at different times $t_i$ close to different field maxima.\\

In conclusion, we have shown that single ionization of hydrogen by a moderately strong ultrashort laser pulse, in the transition regime from multiphoton to tunneling ionization, gives rise to a complex interference pattern in the two-dimensional momentum $(k_z, k_\rho)$ plane. While at high momenta remnants of ATI rings remain visible, at small $k$ near threshold Ramsauer-Townsend diffraction oscillations develop at fixed $k$ as a function the angle between emission direction and laser polarization. A simple semiclassical analysis identifies the fringes resulting from interfering paths released at different times but reaching the same Kepler asymptote. The present result shows that a proper semiclassical description along the lines of the ``simple man's model'' \cite{corkum} requires a three-dimensional description to account for Coulomb scattering. Our results feature a striking similarity to recent data by Rudenko et al. \cite{Rudenko04} suggesting the presence of the 2D interference fringes to be independent of the specific atomic core potential.
		
									The authors acknowledge support by the SFB 016 ADLIS and the project P15025-N08 of the FWF (Austria) and by EU project HPRI-2001-50036, and are grateful to A. Rudenko and J. Ullrich for exchanging data.

%\end{references}

%\newpage

\end{document}